\documentclass[letterpaper,twocolumn,10pt]{article}
\usepackage{usenix-2020-09}
\usepackage{authblk}

\usepackage{cite}
\usepackage{amsmath,amssymb,amsfonts}
\usepackage{algorithmic}
\usepackage{graphicx}
\usepackage{textcomp}
\usepackage{lipsum}
\usepackage{comment}
\usepackage{xurl}
\setlipsum{%
  par-before = \begingroup\color{gray},
  par-after = \endgroup,
  sentence-before = \begingroup\color{gray},
  sentence-after = \endgroup
}
\usepackage{cleveref}
\crefformat{section}{\S#2#1#3} 
\crefformat{subsection}{\S#2#1#3}
\crefformat{subsubsection}{\S#2#1#3}

\crefname{figure}{Fig.}{Figs.}
\crefname{equation}{Eqn.}{Eqns.}
\crefname{section}{\S}{\S}
\crefname{lstlisting}{Listing}{Listings}

\usepackage{tikz}
\usepackage{amsmath}

\newcommand{\insertFigure}[5]{
    \begin{figure}[t]
      \centering
      \includegraphics[width=#3\linewidth]{figure/#1}
      \vspace{#4}
      \caption{\small #2}
      \label{fig:#1}
      \vspace{#5}
    \end{figure}
}

\newcommand{\sysname}{\texttt{G{\small ENIE}}}

\begin{document}
%-------------------------------------------------------------------------------

%don't want date printed
\date{}

\title{\Large \bf Towards Easy and Realistic Network Infrastructure Testing for \\Large-scale Machine Learning}
\author{
Jinsun Yoo\textsuperscript{1}, 
ChonLam Lao\textsuperscript{2}, 
Lianjie Cao\textsuperscript{3}, 
Bob Lantz\textsuperscript{3}, 
Minlan Yu\textsuperscript{2}, 
Tushar Krishna\textsuperscript{1},
Puneet Sharma\textsuperscript{3}
\\
\textsuperscript{1}Georgia Institute of Technology, 
\textsuperscript{2}Harvard University, 
\textsuperscript{3}Hewlett Packard Labs
}
\maketitle

%-------------------------------------------------------------------------------
\begin{abstract}
%-------------------------------------------------------------------------------
This paper lays the foundation for \sysname{}, a testing framework that captures the impact of real hardware network behavior on ML workload performance, without requiring expensive GPUs. \sysname{} uses CPU-initiated traffic over a hardware testbed to emulate GPU to GPU communication, and adapts the ASTRA-sim simulator to model interaction between the network and the ML workload. 
\end{abstract}

%-------------------------------------------------------------------------------
\section{Introduction}
%-------------------------------------------------------------------------------

The growth in both model size and training data has pushed ML training clusters to scale beyond tens of thousands of GPUs~\cite{megascale, metaroce, deepspeed, megatron}. Distributed training involves communication where GPUs periodically share the results of partial computation (such as activations and gradients). 
Communication often becomes a bottleneck for ML training, diminishing the returns from scaling compute. For example, Mixture of Experts (MoE), an increasingly popular technique of Large Language Models, involves all-to-all communication in the critical path~\cite{janus,lina,deepseek}. 

A large GPU cluster relies on an equally complex network infrastructure that spans across multiple NICs, switches, and links. Each node is connected to the network via multiple high-bandwidth NICs that enable low-latency, high-bandwidth communication between GPUs across nodes.
Several layers of network switches connect tens of thousands of such nodes, while links provide the physical connection between switches or switches and NICs. Placing and configuring this set of hardware is a non-trivial task. Correct configuration across NICs and switches are necessary to enable load-balancing schemes like multipathing or spraying, or traffic engineering optimizations that are tailored to the unique characteristics of ML training traffic. Furthermore, the large number of hardware components exposes the network infrastructure to failures or degraded performance, adding to the management and debugging challenges
~\cite{flor, falcon, uecwhitepaper, awssrd, nvidia-adaptive-routing, jupiter2022}.

This complexity motivates a number of use cases for a framework to test the \textit{real network infrastructure}. First, we want to understand how different network configurations (from HW buffer size to congestion control algorithms) or events (such as failures) influence the ML workload
~\cite{megascale, metaroce, topoopt, hpn, CASSINI}.
Specifically, we want to understand how these effects propagate through a workload and impact overall performance. Second, there is vendor demand to validate the network infrastructure before running ML training workloads. Industry operators report that network hardware failures or misconfigurations account for a significant portion of costly failures and restarts~\cite{megascale, metaroce}.

While prior work has modeled network behavior with simulators,
simulators alone are not enough to fully satisfy the above use cases. First, publicly available simulators may not accurately model novel proprietary networks such as HPE Slingshot~\cite{slingshot}. Additionally, simulators cannot validate if the hardware network is configured properly to deliver the desired performance for the ML workload. Finally, simulators cannot easily model unexpected network anomalies that occur in real deployments. Figure~\ref{fig:eval} depicts how the performance of an AllReduce collective suffers as an unpredicted NIC degradation occurs. We repeatedly issue AllReduce collectives across 16 nodes equipped with A100 GPUs and Connectx-6 NICs and plot the performance. The simulator does not reflect the network anomaly that it did not predict.

Unfortunately, to execute an ML workload on a real network infrastructure one must secure an equally large number of GPUs. This is increasingly difficult due to GPU scarcity and cost. We draw attention to the fact that our focus is on the network traffic behavior on the network hardware, not on how computation is done on the GPU. To the best of our knowledge, there is no framework that can execute ML workloads and generate real network traffic without GPUs.

In this paper, we share our vision for \sysname{} (\underline{G}PUs \underline{E}liminated for \underline{N}etwork \underline{I}nfrastructure \underline{E}xamination). \sysname{} provides realistic ML workload performance testing of network infrastructure without 
requiring expensive GPUs. The key idea is to replicate GPU behavior and generate network traffic from the CPU, while capturing the workload behavior through the ASTRA-sim simulator~\cite{astrasim}.

\insertFigure{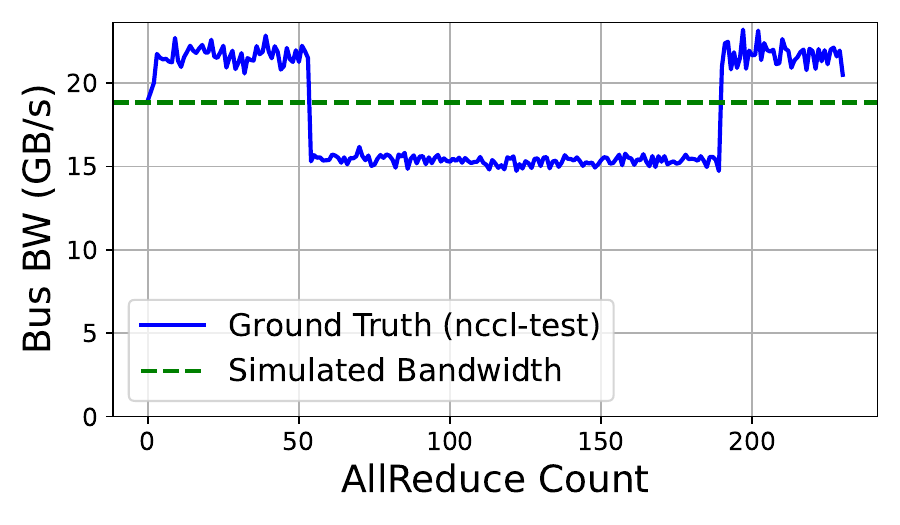}{Bandwidth for several AllReduce runs against simulation. We inject an anomaly which the simulation does not anticipate.}{0.9}{-1.5em}{-1em}
\section{Background}
\subsection{ML Training Process Behavior}

Developers write complex ML training processes with frameworks such as PyTorch~\cite{pytorch} or JAX~\cite{jax}. These processes interact with various hardware components and run across multiple nodes. \cref{fig:training} depicts a typical deployment, where a large number of nodes are connected through a multi-layered network fabric. Within each node, the training process runs on the host CPU, handling operations such as launching compute kernels on accelerator devices (commonly GPUs) and streaming data from host memory to device memory as input to compute kernels. Training nodes must periodically share data stored in memory such as weights, activations, and gradients through collective communication. While the training process initiates the communication, the data bypasses the host OS and is passed into the network directly through transports such as RDMA. 

The dependency of these operations is determined by factors such as the model definition and the parallelization strategies. The training process executes each operation whose upstream operation has completed and the dependency has been resolved. If an operation lies on the critical path, delays can degrade the performance of the overall workload.

\subsection{Communication in ML Training} 

As ML models grow explosively, distributed training increasingly relies on various parallelism schemes to leverage distributed computational resources (e.g., expert parallelism and pipeline parallelism), leading to a high and complex communication overhead.
While this overhead amplifies the need for specialized network designs, the complexity of the network makes management, debugging, and identifying best practices more challenging.

Building and managing a network infrastructure for ML traffic has become increasingly complex. For example, ML traffic requires high-bandwidth, low-latency communication across GPUs, making GPUDirect RDMA and low-level transports (e.g., congestion control) essential. Additionally, ML traffic exhibits unique workload characteristics, such as low-entropy patterns, necessitating load-balancing schemes like multipathing or spraying at the hardware NIC and switches. Furthermore, topology and traffic engineering optimizations tailored to ML workloads add to network complexity. These increasing network demands make management, debugging, and identifying best practices more challenging.

\subsection{Representing ML Workload}

\insertFigure{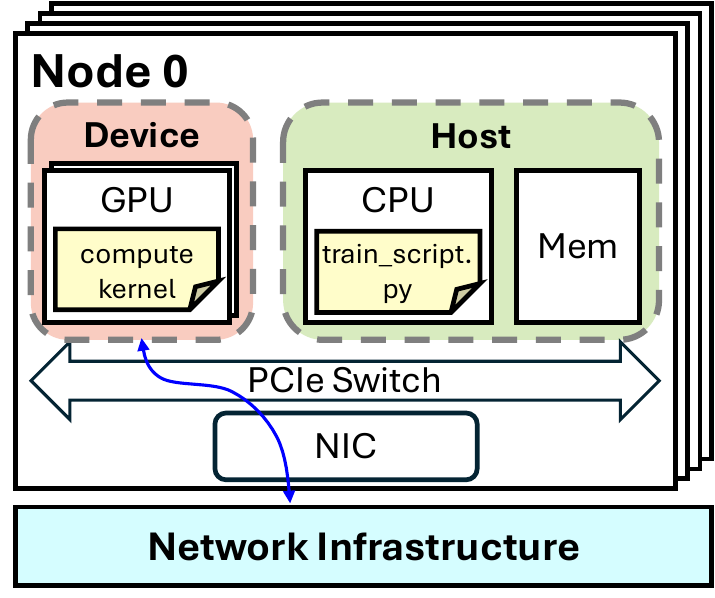}{A typical deployment of training processes across multiple nodes. The blue arrow depicts the communication between nodes.}{0.61}{-1em}{-1.5em}

To study workload behavior without rerunning it on all of the hardware, a suitable representation format is essential. Chakra~\cite{chakra}, supported by the MLCommons initiative~\cite{chakra-wg-page}, is a widely used graph based representation of AI/ML workloads. It captures the execution of distributed workloads into a graph, where vertices denote operators and edges denote their dependencies.

Several tools take a Chakra graph as input, allowing us to easily study a wide array of \textit{arbitrary} workloads. One of such tools is the ASTRA-sim distributed ML simulator~\cite{astrasim}. ASTRA-sim is an event-based simulator. It uses a modular design where users choose between different compute, memory, or communication models. For example, ASTRA-sim supports a network backend based on the ns-3 simulator, which allows it to simulate RDMA traffic at packet level using ns-3 features~\cite{astrasimns3}.
\section{Motivation}
\label{sec:motivation}
In this section, we describe the motivation behind \sysname{} in detail, and discuss how existing work falls short of our goal.

\subsection{Studying the Impact of Network on AI/ML Workload}
\label{subsec:capture}
When building the network for AI/ML workloads, it is crucial to understand how the network interacts with the workload. Researchers seek to understand important questions such as \textit{What is the best set of network configuration or hardware topology I can use for a given workload?} or, \textit{I developed a new congestion control mechanism. How can I showcase its efficiency for AI/ML workloads?}. Answering these questions helps improve system design or allows us to more effectively showcase and validate new ideas.

A simple way to evaluate the above questions is to prepare a training cluster and run the training workload on it. However, this does not scale very well due to the expensive cost of GPUs. 
As a result, academic evaluations have been limited to a small set of clusters. Even in large industry participants, non-production research projects are limited to small evaluations.

An alternative direction would be to use a simulator. For example, some works use the ns-3 simulator to simulate RDMA messages~\cite{astrasimns3, simai}. However, simulators can only model publicly available protocols such as RDMA over Converged Ethernet (RoCE)~\cite{roce}. They are limited in modeling commonly used proprietary technology, such as Infiniband, Slingshot, or Spectrum-X~\cite{infiniband, slingshot, spectrumx}. Another inherent limitation of simulators is their inaccuracy compared to real deployed hardware and software.

One deficiency \sysname{} aspires to solve is the ability to capture the relation between network events and the whole workload, not just a single collective. Simply testing how collectives perform on different configurations is not enough. The unique characteristics of ML traffic such as bursty behavior means that single collective benchmarks may not accurately reflect the statistic pattern of whole workload performance.

For this reason, benchmarking tools like nccl-tests~\cite{nccl-tests}, which only measure a single collective operation across different network configurations, are not suitable for our goals (not to mention that they require GPUs). Gloo-benchmark~\cite{gloo} can run on CPUs without GPUs, but it is also limited to a single collective and is not a suitable option for studying the network-workload relationship.

When running a workload with a network testing framework, we want to capture the interaction between the workload and network components. Some benchmarking tools in other domains trace workloads and replay operations at fixed times based on the original issued timestamps~\cite{rocksdb}. However, operations within ML workloads have dependencies between each other, and simply replaying them at fixed timestamps fails to capture the true workload behavior.

\subsection{Easily Verifying Network Integrity}
Some hardware failure or software misconfigurations could be diagnosed before starting a training job, saving precious GPU hours. For example, Meta reports their experience of diagnosing bad performance in production training jobs, which was tracked back to a mismatch in the packet scheduling algorithms loaded on different switches~\cite{metaroce}. In the end, they had to involve the network vendor to fix the issue. Similarly, while developing \sysname{}, we found that congestion control was disabled in our network switch, which led to backpressure and packet drops. Both of these examples could have been identified before training began. 

However, diagnosing these issues are difficult. Inspecting every hardware/software component is tedious in a largescale deployment. On the other hand, it is expensive for network vendors to purchase large numbers of GPUs for the sole purpose of testing their network infrastructure. Simulators can help (to some extent) predict how failures affect workload performance. However, they cannot detect failures themselves. Altogether, this motivates the case for a network testing framework that does not rely on costly GPUs.

\section{\sysname{} Design}

\insertFigure{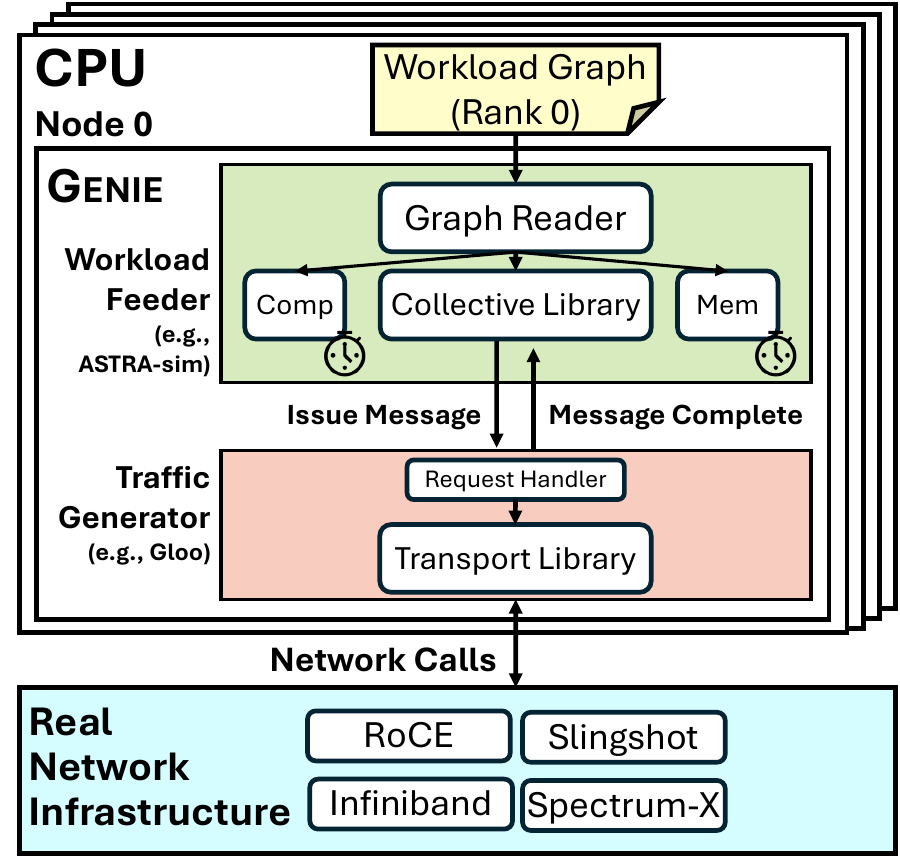}{Overall Architecture of \sysname{}. \sysname{} can work on a cluster with only CPU nodes connected via the network infrastructure.}{0.8}{-1.2em}{-1.5em}

Figure~\ref{fig:arch} is an overview of our design of \sysname{}. \sysname{} has two key components: the \textbf{traffic generator} generates network traffic from CPU\footnote{Intra-node communication is beyond the scope of this work.}, while the \textbf{workload feeder} captures the interaction between the workload and the real network.

\textbf{Modeling Workload with ASTRA-sim:} 
\sysname{} uses the ASTRA-sim simulator to capture the workload behavior of real training processes.
By design, each instance of \sysname{} replicates the behavior of the actual training process. The default ASTRA-sim is a sequential process that simulates a single event based timeline. We modify the simulator to run in a distributed setting. \sysname{} instances are duplicated across CPU nodes, where each instance represents a training rank. \sysname{} uses ASTRA-sim's graph reader to traverse through the workload graph provided as input, which encodes operators and their dependencies. The graph reader issues each operator as discussed below. Once an operator has finished, the graph reader then issues the operators whose dependencies are resolved.

ASTRA-sim models non-communication operators locally and \texttt{sleep}s for the simulated duration as opposed to launching GPU kernels. For collectives, its internal collective library breaks down a collective into send and receive messages. The collective library triggers the traffic generator to issue the messages, and the traffic generator reports once the message finishes. If the collective completes, the graph reader moves on to the next operation. 

ASTRA-sim's collective library allows users to test common collective algorithms such as Ring or Tree. Alternatively, it also allows users to use arbitrary collective algorithms generated by synthesizers or Domain Specific Languages~\cite{taccl, tacos, mscclang, collectiveapi}. 

Note how we are not actually training a model with real inputs and weights. Instead, we simply recreate the local delay a real training process would see for compute or memory operations. While we do generate real RDMA traffic with the same message size as in the training process, the payload contains meaningless data.

The distributed instances of \sysname{} do not use a special synchronization mechanism. Instead, they communicate through collective communication, just as training processes do on a GPU cluster. Each instance emulates compute and memory operations locally. They communicate over the network fabric only when a collective operation occurs in the original workload. 

\sysname{} captures the relationship between network behavior and the workload performance as follows: An optimal network leads to reduced time in the collective communication (and vice versa for a suboptimal network configuration). If there are dependent operations, the graph reader stops traversing the workload graph and waits for the collective dependency completes and the dependency is resolved. In the end, any delay along the critical path slows down the workload process. As a result, \sysname{} can capture how network issues impact workload performance.

\textbf{Creating GPU Communication with Traffic Generator:}
We implement a traffic generator as the network backend for ASTRA-sim. A key requirement of \sysname{} is to generate GPU traffic with CPUs. The traffic generator acts as an interface between ASTRA-sim and the real network. It exposes endpoints for point-to-point send and receive messages. Once ASTRA-sim's collective library breaks down a collective into send and receive messages, it calls the traffic generator through these endpoints. The traffic generator then creates and injects traffic into the network.

Once ASTRA-sim triggers the traffic generator, the request handler calls the transport library to perform tasks such as assigning memory buffers or issuing network calls. To generate RDMA traffic, for example, the traffic generator can directly call libraries such as libibverbs~\cite{libibverbs} or libfabric~\cite{libfabric}. Alternatively, it can also use high level libraries that encapsulate the low level libraries. Examples include the point-to-point messages functions of Gloo~\cite{gloo} or perftest~\cite{perftest}. Both tools can generate arbitrary point-to-point messages only with CPUs in an application agonistic manner. Choosing the right implementation to balance fidelity and flexibility is left to future work.

While we discuss generating RDMA traffic as an example, \sysname{} is designed to be easily extensible. \sysname{} is envisioned to work on arbitrary network infrastructure supporting various transports. The traffic generator should switch between different software stack or network drivers depending on the network fabric it is deployed on. 

\textbf{Network Infrastructure}
The direct interaction with real network is what sets \sysname{} apart from simulators. Once the traffic generator injects network traffic into the network through the NIC, the traffic traverses through the underlying network fabric to the destination. The switches in the network fabric do not differentiate traffic generated with \sysname{} from traffic generated from training processes on GPU clusters.

Injecting traffic directly into the real network fabric allows \sysname{} to realistically and accurately test network hardware and validate different configurations. \sysname{} is designed to be modular and portable in nature. This will allow us to deploy and test \sysname{} on a wide array of production networks, such as RoCE, Slingshot, Infiniband, and Spectrum-X.
 
\section{Conclusion and Future Work}
In this paper we motivated \sysname{}, a framework to test real network infrastructure for large-scale ML workloads without requiring costly GPUs. 
Several efforts remain to fully realize the vision of \sysname{}. First we will build \sysname{} and carefully study and validate the fidelity of the GPU emulation. We will then run evaluations across a wide range of network configurations, workloads, and failure scenarios showcasing \sysname{}'s  potential.

%-------------------------------------------------------------------------------
\bibliographystyle{plain}
\bibliography{reference.bib}

%%%%%%%%%%%%%%%%%%%%%%%%%%%%%%%%%%%%%%%%%%%%%%%%%%%%%%%%%%%%%%%%%%%%%%%%%%%%%%%%
\end{document}